\documentclass[a4paper, 10pt, conference]{cssconf}
\IEEEoverridecommandlockouts    
\usepackage{epsfig} 
\usepackage{amsmath} 
\usepackage[misc]{ifsym}

\title{\LARGE \bf
A Grasping Movement Intention Estimator \\ 
for Intuitive Control of Assistive Devices}

\author{Etienne Moullet $^{\textrm{\Letter}}$, Justin Carpentier, Christine Azevedo-Coste and François Bailly
\thanks{E. Moullet is with WILLOW, INRIA - \'Ecole Normale Supérieure - PSL Research University, Paris, France(corresponding author, e-mail: {\tt\small etienne.moullet@inria.fr}).}
\thanks{J. Carpentier is with WILLOW, INRIA - \'Ecole Normale Supérieure - PSL Research University, Paris, France.}
\thanks{C. Azevedo-Coste is with CAMIN, INRIA centre d'Université Côte d'Azur, Université de Montpellier, Montpellier, France.}
\thanks{F. Bailly is with CAMIN, INRIA centre d'Université Côte d'Azur, Université de Montpellier, Montpellier, France.}
}

\begin{document}
\maketitle
\thispagestyle{empty}
\pagestyle{empty}

\begin{abstract}

This study introduces i-GRIP, an innovative movement goal estimator designed to facilitate the control of assistive devices for grasping tasks in individuals with upper-limb impairments. The algorithm operates within a collaborative control paradigm, eliminating the need for specific user actions apart from naturally moving their hand toward a desired object. i-GRIP analyzes the hand's movement in an object-populated scene to determine its target and select an appropriate grip. In an experimental study involving 11 healthy participants, i-GRIP exhibited promising estimation performances (success rates of 89.9\% for target identification and 94.8\% for grip selection) and responsiveness (mean delays of 0.53s for target identification and 0.39s for grip selection), showing its potential to facilitate the daily use of grasping assistive devices for individuals with upper-limb impairments.

\end{abstract}

\section{INTRODUCTION}

Upper-limb impairments, such as spinal cord injury, stroke, or amputation, can significantly impact an individual's quality of life and autonomy. These impairments can make it difficult or impossible to perform many daily activities that require grasping, such as eating, dressing, and grooming. Various approaches and assistive devices have been developed to compensate for these difficulties, including functional electrical stimulation (FES), exoskeletons, and prostheses. However, despite continuous improvements in terms of actuation speed, accuracy, and strength \cite{gantenbein}, the user is often hindered in the task execution by these devices' control modalities (manipulating joystick/switch buttons, detection of neural activity, muscular activation or stereotypical movements, voice recognition etc. \cite{azevedo, jiang_farina}).

    For instance, individuals with quadriplegia but with residual upper-limb motion (typically able to move their arms but not their fingers) have a control over their body too reduced to efficiently express their grasping intent, and current human-machine interfaces provide limited inputs for controlling devices \cite{jiang_farina}. Current approaches thus often rely on state machines to alternatively set a given user's action as the control input for a given movement elicited by the device. However, this requires the user to constantly switch between modes to achieve daily tasks, which can be cognitively demanding, result in saccadic movements, and, most importantly, impact the user's ability to perform the task effectively.

In this study, we present i-GRIP, a novel grasping movement intention estimator. i-GRIP operates within a collaborative control paradigm, requiring no specific action from the users except naturally moving their hand toward the object they wish to grasp. The algorithm performs a kinematic analysis of the hand's movement in the observed scene to identify the targeted object and an appropriate grip to grasp it. These information can then be used to control upper-limb grasping assistive devices.

\section{Methods and materials}

\subsection{\scshape i-GRIP algorithm}
\begin{figure}
        \centering
        \includegraphics[width = 0.9\linewidth]{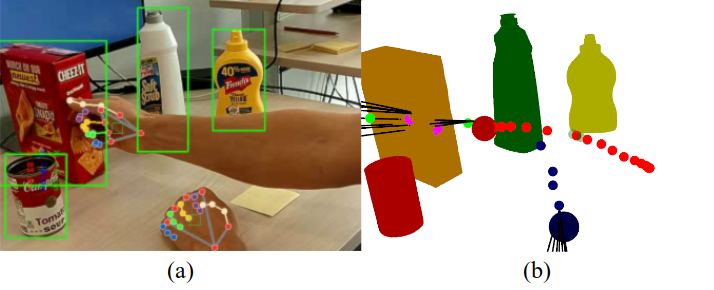}
        \caption{Experimental setup and pre-processing: (a) - Example of a video frame captured during a trial overlaid with green rectangles marking the detected objects and multicolored landmarks marking hands keypoints. (b) - Corresponding 3D virtual scene: Orange, yellow, red, and green objects are the rendered meshes of the detected objects. Big blue and red spheres represent the 3D positions of, respectively, left and right hands. The middle-sized red and green spheres represent, respectively, the past and expected future trajectory of the right hand. The black lines are a cone of rays expanding from the expected future trajectory of the right hand, and whose impacts are the small magenta dots on the mesh. }
        \label{fig:setup}
    \end{figure}

{\scshape i-GRIP} is designed to work downstream of any scene observation process able to provide hands 3D position and objects 6D poses (see subsection \ref{implem} for a possible implementation). Detected hands and objects are represented in a virtual 3D duplicate scene (see Fig. \ref{fig:setup}-b). At each time step and new measurement, the algorithm performs a kinematic analysis of the hands' motions to predict their near-future trajectories. Then, for each hand-object pair, confidence scores are computed based on four metrics that describe the motion of a hand $\boldsymbol{h}$ relatively to the object $\boldsymbol{o}^j$:
\begin{itemize}
    \item \textbf{ray impacts:} the number of impacts onto the object's mesh from cones of rays, cast from near-future trajectory points in the direction of the local velocity vector,
    \item \textbf{distance derivative:} the time derivative of the distance between the hand's position and the position of the center of the object's mesh,
    \item \textbf{distance:} the distance between the hand's position and the object's mesh,
    \item \textbf{future distance:} the distance between the barycenter of near-future trajectory points and the object's mesh.
\end{itemize}
The cone of rays is intended to handle curved, fast trajectory parts, while distance derivative would be more efficient on straight, fast trajectory parts. On the other hand, distance and future distance aim at slower hand displacements. Together, these metrics may handle diverse trajectories at all stages.

Next, a velocity-dependent weighted sum of these 4 confidence scores is performed, defining a global confidence score $c_{glob}(\boldsymbol{h}|\boldsymbol{o}^j)$.
Then, the target of the hand's movement is identified as the observed object with the highest confidence score:  
        \begin{equation}\label{eq_target_identification}
            target(\boldsymbol{h}) = \underset{\boldsymbol{o}^j}{\arg \max}(c_{glob}( \boldsymbol{h}| \boldsymbol{o}^j))
        \end{equation}

\begin{figure}
\centering
\includegraphics[width=0.6\linewidth]{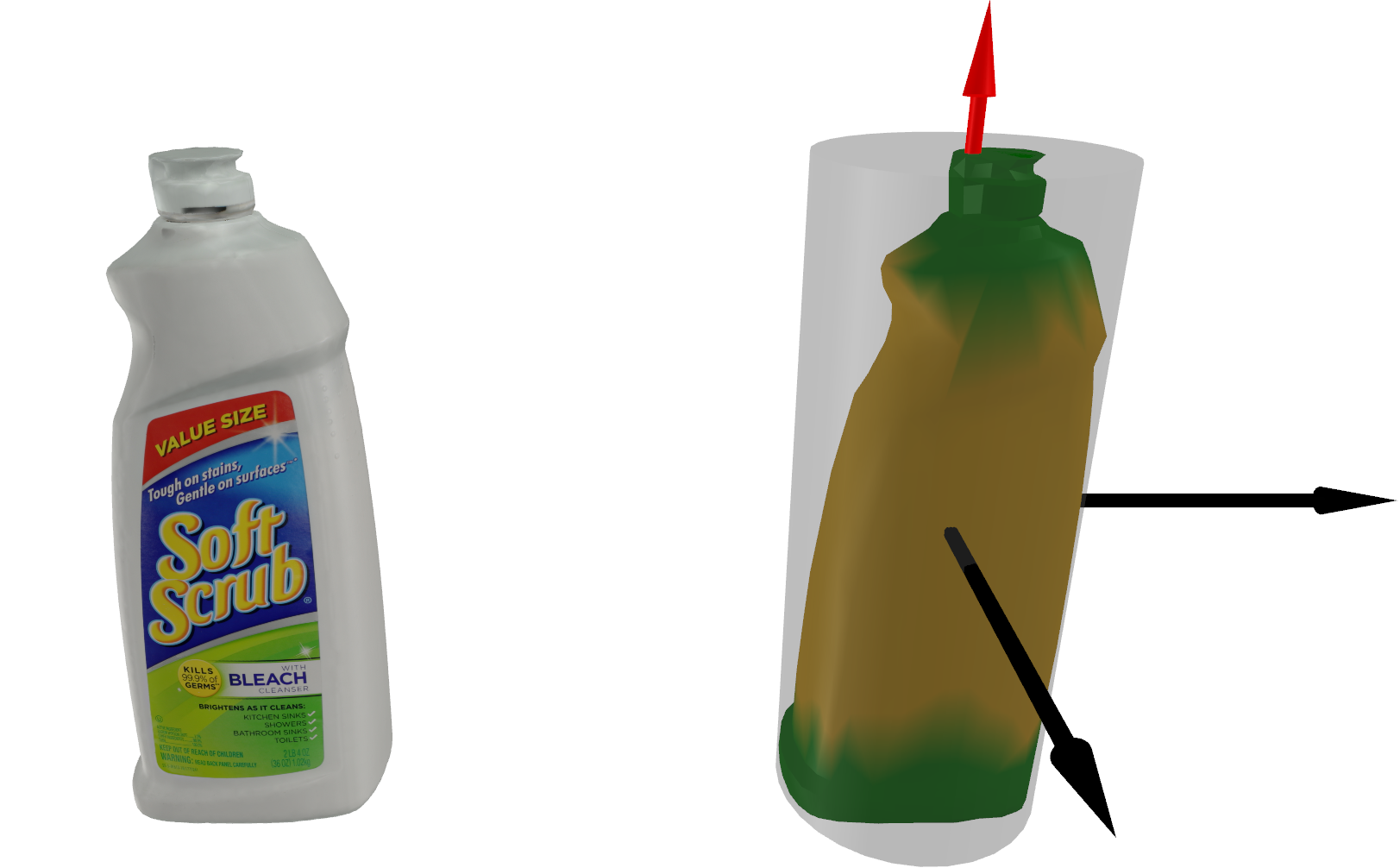}
 \caption{Bleach bottle from YCB set \cite{ycb} (left) and a visualization of the corresponding grip selection process (right). Transparent grey volume is the bounding cylinder of the mesh. Red arrow figures the z-axis of the object’s frame. The yellow zone illustrates the z-values corresponding to a palmar grip, while the green zone and outwards correspond to a pinch grip.}
\label{fig:grip}
\end{figure}

\noindent Finally, the appropriate grip is determined by the hand's position relative to the detected objects and their shapes.
In its current version, i-GRIP focuses on oblong objects that may be grasped using two main grips used in daily life \cite{feix}: palmar and pinch grips.
First, the mesh of each detected object $\boldsymbol{o}^j$ is bounded by a cylinder. Its axis of revolution defines the z-axis of the object's reference frame $R^j$ whose origin is placed at the center of gravity of the mesh. 
Then, a grip for each hand-object pair $grip(\boldsymbol{h}|\boldsymbol{o}^j)$ is determined by the z-component of the hand's position in $R^j$ (see Fig. \ref{fig:grip}). Finally, the grip appropriate for the analyzed movement is the grip  corresponding to the target found with \eqref{eq_target_identification}:
        \begin{equation}
             grip(\boldsymbol{h})=grip(\boldsymbol{h}| target(\boldsymbol{h}))
        \end{equation}

\subsection{Experimental study}\label{implem}
An experimental study approved by INRIA ethical committee (COERLE Decision 2024-01) involving eleven healthy participants was conducted to evaluate i-GRIP's performance. Participants were seated in front of a table (see Fig. \ref{fig:setup}-a) and performed 128 grasping movements under homogeneously drawn conditions among the following:
        \begin{itemize}
            \item which hand to use: left or right,
            \item which objects to target among 4 objects from the YCB set \cite{ycb} placed on the table: a mustard bottle, a bleach bottle, a tomato can or a box of cheez'it,
            \item which grip to apply: pinch or palmar,
            \item whether to execute the grip (actually grasping the object) or simulate it (not moving their fingers during movement nor grasping the object).
        \end{itemize}
Two stereoscopic cameras (OAK-D S2, Luxonis) were placed on the left and right sides of participants at shoulder level and filmed the whole scene (hands and the four objects on the table). RGB frames and depth maps were recorded and processed offline with computer vision tools (that are not part of i-GRIP, but upstream of it) to extract the hands' and objects' observations i-GRIP takes as inputs. Hands 3D positions were estimated using mediapipe~\cite{mediapipe} and depth maps. Objects 6D poses were estimated using a version of CosyPose \cite{cosypose} trained on the YCB dataset \cite{ycb}. Video trials for which hands or object detections were not successful enough were excluded from the study. Target identification and grip selection were performed on every trial's frames and compared to the instructions ground truth. {\scshape i-GRIP} was deemed successful over a trial when it was successful over more than 70\% of its frames.

\section{Results}
{\scshape i-GRIP} successfully identified the target in 89.9\% of the recorded movements and selected the correct grip in 94.8\% of them. Targets were identified within a mean delay of 0.52s and grips within a mean delay of 0.39s, leaving mean temporal margins before the end of the movement of, respectively, 0.67s and 0.80s.

\section{Conclusion}
{\scshape i-GRIP} effectively identified the target of grasping movement and selected appropriate grips within less than half the mean duration of movements, regardless of the experimental conditions (camera placement and movement type). Further studies must assess i-GRIP's effectiveness with different observation devices, in real-life scenarios, pathologies, and assistive devices and possibly fine-tune its parameters accordingly.



\end{document}